\input boxedepsf.tex 
\SetRokickiEPSFSpecial%

 \def\LOADED{\relax}
 \let\@temp=\relax
 \ifx\BoxedEPSFLoaded\LOADED
  \immediate\write16{}
  \immediate\write16{  BoxedEPSF macros already defined!}
  \immediate\write16{}
  \let\@temp=\endinput
  \fi\@temp

 \let\BoxedEPSFLoaded\LOADED

 \chardef\CatAt\the\catcode`\@
 \chardef\CatColon\the\catcode`\:
 \chardef\CatDollar\the\catcode`\$
 \catcode`\@=11
 \catcode`\:=12

 \let\wlog@ld\wlog
 \def\wlog#1{\relax}

 \newif\ifIN@
 \newdimen\XShift@ \newdimen\YShift@
 \newtoks\Realtoks

 \newdimen\Wd@ \newdimen\Ht@
 \newdimen\Wd@@ \newdimen\Ht@@
 \newdimen\TT@
 \newdimen\LT@
 \newdimen\BT@
 \newdimen\RT@
 \newdimen\XSlide@ \newdimen\YSlide@
 \newdimen\TheScale  
 \newdimen\FigScale  
 \newdimen\ForcedDim@@

 \newtoks\EPSFDirectorytoks@
 \newtoks\EPSFNametoks@
 \newtoks\BdBoxtoks@

 \newif\ifNotIn@
 \newif\ifForcedDim@
 \newif\ifForcedHeight@
 \newif\ifPSOrigin

 \newread\EPSFile@

 \newif\ifIN@\def\IN@{\expandafter\INN@\expandafter}
  \long\def\INN@0#1@#2@{\long\def\NI@##1#1##2##3\ENDNI@
    {\ifx\m@rker##2\IN@false\else\IN@true\fi}%
     \expandafter\NI@#2@@#1\m@rker\ENDNI@}
  \def\m@rker{\m@@rker}

  \newtoks\Initialtoks@  \newtoks\Terminaltoks@
  \def\SPLIT@{\expandafter\SPLITT@\expandafter}
  \def\SPLITT@0#1@#2@{\def\TTILPS@##1#1##2@{%
     \Initialtoks@{##1}\Terminaltoks@{##2}}\expandafter\TTILPS@#2@}

  \newtoks\Trimtoks@

 \def\ForeTrim@{\expandafter\ForeTrim@@\expandafter}
 \def\ForePrim@0 #1@{\Trimtoks@{#1}}
 \def\ForeTrim@@0#1@{\IN@0\m@rker. @\m@rker.#1@%
     \ifIN@\ForePrim@0#1@%
     \else\Trimtoks@\expandafter{#1}\fi}

  \def\Trim@0#1@{%
      \ForeTrim@0#1@%
      \IN@0 @\the\Trimtoks@ @%
        \ifIN@
             \SPLIT@0 @\the\Trimtoks@ @\Trimtoks@\Initialtoks@
             \IN@0\the\Terminaltoks@ @ @%
                 \ifIN@
                 \else \Trimtoks@ {FigNameWithSpace}%
                 \fi
        \fi
      }

   \newtoks\pt@ks
   \def \getpt@ks 0.0#1@{\pt@ks{#1}}
  \newtoks\Realtoks
  \def\Real#1{%
    \dimen2=#1%
      \SPLIT@0\the\pt@ks @\the\dimen2@
       \Realtoks=\Initialtoks@
            }

   \newdimen\Product
   \def\Mult#1#2{%
     \dimen4=#1\relax
     \dimen6=#2%
     \Real{\dimen4}%
     \Product=\the\Realtoks\dimen6%
        }

 \newdimen\Inverse
 \newdimen\hmxdim@ \hmxdim@=8192pt
 \def\Invert#1{%
  \Inverse=\hmxdim@
  \dimen0=#1%
  \divide\Inverse \dimen0%
  \multiply\Inverse 8}

   \def\Rescale#1#2#3{
              \divide #1 by 100\relax
              \dimen2=#3\divide\dimen2 by 100 \Invert{\dimen2}%
              \Mult{#1}{#2}%
              \Mult\Product\Inverse
              #1=\Product}

  \def\Scale#1{\dimen0=\TheScale %
      \divide #1 by  1280 
      \divide \dimen0 by 5120 %
      \multiply#1 by \dimen0
      \divide#1 by 10   
     }

 \newbox\scrunchbox

 \def\Scrunched#1{{\setbox\scrunchbox\hbox{#1}%
   \wd\scrunchbox=0pt
   \ht\scrunchbox=0pt
   \dp\scrunchbox=0pt
   \box\scrunchbox}}

 \def\Shifted@#1{%
   \vbox {\kern-\YShift@
       \hbox {\kern\XShift@\hbox{#1}\kern-\XShift@}%
           \kern\YShift@}}

 \def\cBoxedEPSF#1{{}\leavevmode 
   \ReadNameAndScale@{#1}%
   \SetEPSFSpec@
   \ReadEPSFile@ \ReadBdB@x
     \TrimFigDims@
     \CalculateFigScale@
     \ScaleFigDims@
     \SetInkShift@
   \hbox{$\mathsurround=0pt\relax
         \vcenter{\hbox{%
             \FrameSpider{\hskip-.4pt\vrule}%
             \vbox to \Ht@{\offinterlineskip\parindent=\z@%
                \FrameSpider{\vskip-.4pt\hrule}\vfil
                \hbox to \Wd@{\hfil}%
                \vfil
                \InkShift@{\EPSFSpecial{\EPSFSpec@}{\FigSc@leReal}}%
             \FrameSpider{\hrule\vskip-.4pt}}%
         \FrameSpider{\vrule\hskip-.4pt}}}%
     $}%
    \CleanRegisters@
    }

 \def\tBoxedEPSF#1{\setbox4\hbox{\cBoxedEPSF{#1}}%
     \setbox4\hbox{\raise -\ht4 \hbox{\box4}}%
     \box4
      }

 \def\bBoxedEPSF#1{\setbox4\hbox{\cBoxedEPSF{#1}}%
     \setbox4\hbox{\raise \dp4 \hbox{\box4}}%
     \box4
      }

  \let\BoxedEPSF\cBoxedEPSF

  \def\gLinefigure[#1scaled#2]_#3{%
        \BoxedEPSF{#3 scaled #2}}

  \def\EPSFxsize{\afterassignment\ForceW@\ForcedDim@@}
      \def\ForceW@{\ForcedDim@true\ForcedHeight@false}

  \def\EPSFysize{\afterassignment\ForceH@\ForcedDim@@}
      \def\ForceH@{\ForcedDim@true\ForcedHeight@true}

 \def\ReadNameAndScale@#1{\IN@0 scaled@#1@
   \ifIN@\ReadNameAndScale@@0#1@%
   \else \ReadNameAndScale@@0#1 scaled\DefaultMilScale @
   \fi}

 \def\ReadNameAndScale@@0#1scaled#2@{
    \Trim@0#1@%
    \EPSFNametoks@\expandafter{\the\Trimtoks@}%
    \FigScale=#2 pt%
     }

 \def\SetDefaultEPSFScale#1{%
      \global\def\DefaultMilScale{#1}}

 \SetDefaultEPSFScale{1000}

 \def \SetBogusBbox@{%
     \global\BdBoxtoks@{ BoundingBox:0 0 595 842 }%
     \global\def\BdBoxLine@{ BoundingBox:0 0 595 842 }%
     }

 \def\ReadEPSFile@{
     \openin\EPSFile@\EPSFSpec@
     \relax  
  \ifeof\EPSFile@
     \SetBogusBbox@
     \immediate\write16{}%
     \message{ *** EPS FILE  }%
     \message\expandafter{\the\EPSFNametoks@}%
     \message{ NOT FOUND!  }%
     \immediate\write16{}\relax%
  \else
   \begingroup
   \catcode`\%=12\catcode`\:=12\catcode`\\=12
   \NotIn@true
    \loop
      \ifeof\EPSFile@\NotIn@false
        \SetBogusBbox@
        \immediate\write16{}%
        \message{ *** BoundingBox not found in }%
        \message\expandafter{\the\EPSFNametoks@\space *** }%
        \immediate\write16{}%
      \else\global\read\EPSFile@ to \BdBoxLine@
      \fi
      \global\BdBoxtoks@\expandafter{\BdBoxLine@}%
      \IN@0BoundingBox:@\the\BdBoxtoks@ @%
      \ifIN@\NotIn@false\fi%
    \ifNotIn@\repeat
   \endgroup\relax
  \fi
  \closein\EPSFile@
   }

  \def\ReadBdB@x{
   \expandafter\ReadBdB@x@\the\BdBoxtoks@ @}

  \def\ReadBdB@x@#1BoundingBox:#2@{
    \ForeTrim@0#2@%
    \expandafter\ReadBdB@x@@\the\Trimtoks@ @%
   }

  \newtoks\LLXtoks@  
  \newtoks\LLYtoks@

  \def\ReadBdB@x@@#1 #2 #3 #4@{
      \Wd@=#3bp\advance\Wd@ by -#1bp%
      \Ht@=#4bp\advance\Ht@ by-#2bp%
       \Wd@@=\Wd@ \Ht@@=\Ht@ 
       \LLXtoks@={#1}\LLYtoks@={#2}
      \ifPSOrigin\XShift@=-#1bp\YShift@=-#2bp\fi
     }

   \def\G@bbl@#1{}
   \bgroup
     \global\edef\OtherB@ckslash{\expandafter\G@bbl@\string\\}
   \egroup

  \def\SetEPSFDirectory{
           \bgroup\catcode`\:=12\relax
           \catcode`\$=12\relax
           \SetEPSFDirectory@}

 \def\SetEPSFDirectory@#1{
    \Trim@0#1@
    \global\EPSFDirectorytoks@\expandafter{\the\Trimtoks@ }\relax
    \egroup}

 \def\SetEPSFSpec@{%
     \bgroup
     \let\\=\OtherB@ckslash
     \global\edef\EPSFSpec@{\the\EPSFDirectorytoks@\the\EPSFNametoks@}%
     \global\edef\EPSFSpec@{\EPSFSpec@}%
     \egroup}

 \def\TrimTop#1{\advance\TT@ by #1}
 \def\TrimLeft#1{\advance\LT@ by #1}
 \def\TrimBottom#1{\advance\BT@ by #1}
 \def\TrimRight#1{\advance\RT@ by #1}

 \def\TrimFigDims@{%
    \advance\Wd@ by -\LT@
    \advance\Wd@ by -\RT@ \RT@=\z@
    \advance\Ht@ by -\TT@ \TT@=\z@
    \advance\Ht@ by -\BT@
    }

  \def\ForceWidth#1{\ForcedDim@true
       \ForcedDim@@#1\ForcedHeight@false}

  \def\ForceHeight#1{\ForcedDim@true
       \ForcedDim@@=#1\ForcedHeight@true}

  \def\epsfxsize{\afterassignment\ForceW@\ForcedDim@@}
      \def\ForceW@{\ForcedDim@true\ForcedHeight@false}

  \def\epsfysize{\afterassignment\ForceH@\ForcedDim@@}
      \def\ForceH@{\ForcedDim@true\ForcedHeight@true}

  \def\CalculateFigScale@{%
     \ifForcedDim@\FigScale=1000pt
           \ifForcedHeight@
                \Rescale\FigScale\ForcedDim@@\Ht@
           \else
                \Rescale\FigScale\ForcedDim@@\Wd@
           \fi
     \fi
     \Real{\FigScale}%
     \edef\FigSc@leReal{\the\Realtoks}%
     }

  \def\ScaleFigDims@{\TheScale=\FigScale
      \ifForcedDim@
           \ifForcedHeight@ \Ht@=\ForcedDim@@  \Scale\Wd@
           \else \Wd@=\ForcedDim@@ \Scale\Ht@
           \fi
      \else \Scale\Wd@\Scale\Ht@
      \fi
      \ForcedDim@false
      \Scale\LT@\Scale\BT@  
      \Scale\XShift@\Scale\YShift@
      }

 \def\ShowReservedBoxes{\gdef\FrameSpider##1{##1}}
 \let\HideDisplacementBoxes\HideReservedBoxes  
 
  \ShowReservedBoxes

 \def\hSlide#1{\advance\XSlide@ by #1}
 \def\vSlide#1{\advance\YSlide@ by #1}

  \def\SetInkShift@{%
            \advance\XShift@ by -\LT@
            \advance\XShift@ by \XSlide@
            \advance\YShift@ by -\BT@
            \advance\YShift@ by -\YSlide@
             }
  \def\InkShift@#1{\Shifted@{\Scrunched{#1}}}

  \def\CleanRegisters@{%
      \globaldefs=1\relax
        \XShift@=\z@\YShift@=\z@\XSlide@=\z@\YSlide@=\z@
        \TT@=\z@\LT@=\z@\BT@=\z@\RT@=\z@
      \globaldefs=0\relax}

 \def\SetTexturesEPSFSpecial{\PSOriginfalse
  \gdef\EPSFSpecial##1##2{\relax
    \edef\specialthis{##2}%
    \SPLIT@0.@\specialthis.@\relax
    \special{illustration ##1 scaled
                        \the\Initialtoks@}}}

  \def\SetUnixCoopEPSFSpecial{\PSOrigintrue 
   \gdef\EPSFSpecial##1##2{%
      \dimen4=##2pt
      \divide\dimen4 by 1000\relax
      \Real{\dimen4}
      \edef\Aux@{\the\Realtoks}%
      \includegraphics{##1\space}}}

  \def\SetBechtolsheimRokickiEPSFSpecial{\PSOrigintrue
   \gdef\EPSFSpecial##1##2{%
      \dimen4=##2pt
      \divide\dimen4 by 1000\relax
      \Real{\dimen4}
      \edef\Aux@{\the\Realtoks}%
      \special{ps: psfiginit}%
      \special{ps: literal 1 1 0 0 1 1 startTexFig
           \the\mag\space 1000 div \Aux@\space mul
           \the\mag\space 1000 div \Aux@\space mul scale}%
      \special{ps: include  ##1}%
      \special{ps: literal endTexFig}%
        }}

  \def\SetLisEPSFSpecial{\PSOrigintrue
   \gdef\EPSFSpecial##1##2{%
      \dimen4=##2pt
      \divide\dimen4 by 1000\relax
      \Real{\dimen4}
      \edef\Aux@{\the\Realtoks}%
      \special{pstext="1 1 0 0 1 1 startTexFig\space
           \the\mag\space 1000 div \Aux@\space mul
           \the\mag\space 1000 div \Aux@\space mul scale}%
      \includegraphics{##1}%
      \special{pstext=endTexFig}%
        }}

  \def\SetRokickiEPSFSpecial{\PSOrigintrue
   \gdef\EPSFSpecial##1##2{%
      \dimen4=##2pt
      \divide\dimen4 by 10\relax
      \Real{\dimen4}
      \edef\Aux@{\the\Realtoks}%
      \includegraphics{##1}}}

  \def\SetInlineRokickiEPSFSpecial{\PSOrigintrue
   \gdef\EPSFSpecial##1##2{%
      \dimen4=##2pt
      \divide\dimen4 by 1000\relax
      \Real{\dimen4}
      \edef\Aux@{\the\Realtoks}%
      \special{ps::[begin] 1 1 0 0 1 1 startTexFig\space
           \the\mag\space 1000 div \Aux@\space mul
           \the\mag\space 1000 div \Aux@\space mul scale}%
      \special{ps: plotfile ##1}%
      \special{ps::[end] endTexFig}%
        }}

  \def\SetOzTeXEPSFSpecial{\PSOriginfalse 
  \gdef\EPSFSpecial##1##2{
     \special{##1\space
       ##2 1000 div \the\mag\space 1000 div mul
       ##2 1000 div \the\mag\space 1000 div mul scale
       \the\LLXtoks@\space neg \the\LLYtoks@\space neg translate
             }}}

 \def\SetArborEPSFSpecial{\PSOriginfalse 
   \gdef\EPSFSpecial##1##2{%
     \edef\specialthis{##2}%
     \SPLIT@0.@\specialthis.@\relax 
     \special{ps: epsfile ##1\space \the\Initialtoks@}}}

 \def\SetClarkEPSFSpecial{\PSOriginfalse 
   \gdef\EPSFSpecial##1##2{%
     \Rescale {\Wd@@}{##2pt}{1000pt}%
     \Rescale {\Ht@@}{##2pt}{1000pt}%
     \special{dvitops: import
           ##1\space\the\Wd@@\space\the\Ht@@}}}

 \def\SetStandardEPSFSpecial{%
   \gdef\EPSFSpecial##1##2{%
     \immediate\write16{}
     \immediate\write16{%
       **** Sorry! There is still no standard for \string%
       \special \space EPSF integration *****}%
     \immediate\write16{%
      --- So you will have to identify your driver using a command}%
     \immediate\write16{%
      --- of the form \string\Set...EPSFSpecial, in order to get}%
     \immediate\write16{%
      --- your graphics to print.  See BoxedEPSF.doc.}%
     \immediate\write16{}
     \KillEPSFSpecial
     }}

  \def\KillEPSFSpecial{\gdef\EPSFSpecial##1##2{}}

  \SetStandardEPSFSpecial 

 \let\wlog\wlog@ld 

 \catcode`\@=\CatAt
 \catcode`\:=\CatColon

  \endinput

\documentstyle[12pt,BoxedEPSF]{article}
\HideDisplacementBoxes

\newcommand{\bc}{\begin{center}}
\newcommand{\ec}{\end{center}}
\newcommand{\br}{\begin{flushright}}
\newcommand{\er}{\end{flushright}}
\newcommand{\be}{\begin{eqnarray}}
\newcommand{\ee}{\end{eqnarray}}

\newcounter{tablecounter}
\newcommand{\tablecaption}[1]{
\bc {\bf Table \arabic{tablecounter}.} #1 \\ \ec
\addtocounter{tablecounter}{1}
}
\begin{document}
\pagestyle{empty}

\vbox{\vglue 4cm}
\bc
{\Large \bf Anomalous di-photon production at LEP: \\
    possible consequences at FNAL hadron collider.}
\ec

\vspace{4cm}
\bc{\bf V.A.Litvin,~~~~S.R.Slabospitsky} \\
{\it Institute for High Energy Physics  \\
 Protvino, Moscow Region 142284, RUSSIA}
\vspace{0.5cm}

and

\vspace{0.5cm}
{\bf A.V.Shvorob}  \\
{\it Moscow Institute of Physics and Technology   \\
     Dolgoprudny, Moscow region 141700, RUSSIA},
\ec
\vspace{2cm}
\bc
{\bf Abstract}
\ec
     The hypothetic R--resonance
production (anomalous "$l\bar l\gamma\gamma$" events at LEP)
with 60 GeV mass at FNAL hadron collider has been considered.
The cross-section
of this R-resonance production with the consequent decay
$R\,\to\,\gamma\gamma$ is obtained. Various distributions
of leptons and photons in the process $p\bar p\,\to\,l\bar l
\,R\,(\,\to\,\gamma\gamma)\, X$ are calculated.

\newpage
\bc
\bf INTRODUCTION
\ec
 Nine unusual events of the type
 $e^{+}e^{-}\rightarrow f{\bar f}\gamma\gamma$
with the invariant mass of the two photons near 60 GeV were found from the
statistics collected by the four LEP experiments [1,2,3,4]
(2~--~in $e^+e^-\gamma\gamma$, 5~--~in $\mu^{+}\mu^{-}\gamma\gamma$,
 1~--~in $q{\bar q}\gamma\gamma$,
1~--~in $\nu\bar\nu\gamma\gamma$ channels, respectively
\cite{The Foundations 2}).

 Recent calculations [5,6,7,8] have shown that the Standard Model processes
are unlikely to explain the observed accumulation of the events in a narrow
$M_{\gamma\gamma}$ band near 60 GeV.
 Other feasible
models (including some common extensions of the Standard Model) also
experience considerable difficulties in giving a plausible interpretation
of these events
(see, for example, \cite{other models})

 In \cite{The Foundations 1,The Foundations 2} an attempt to account for
these events was made by assuming
the existence of a scalar (pseudoscalar) resonance $R$, coupling
photons and $Z^0$-bosons only, with its mass close to 60 GeV. In these papers
such a model was shown not to provide a neat explanation of the
(possibly observed) phenomenon, its free parameters being effectively
restricted
by the requirement of the consistency of the model with other experimental
data. Nevertheless, presently available low statistics doesn't let us dismiss
it
confidently and from the phenomenological point of view it appears
interesting to find out if other experiments can add something to the limits
obtained at LEP. Before the advent of new machines (LEP 200, LHC, NLC, etc.)
the hadron collider at FNAL is the most suitable place for such investigations.

 In this paper we analyse the implications of the existence of the
hypothetical resonance for
the physics accessible for
center-of-mass energy and luminosity at the FNAL hadron collider.
The present--day CDF and D0 detectors [13] allow one
to collect some evidence for or against the existence of such resonance.

 The paper is organized as follows. Section 1 describes the
phenomenological model of resonance
interactions with photons and Z--bosons. In Section 2 we
discuss possible manifestations of the resonance in $p\bar p$
collisions and present calculations results.
The results obtained are summarised in Conclusion.

\section{\bf THE PHENOMENOLOGICAL MODEL.}

 As has been already mentioned in Introduction we assume the
existence of a narrow scalar (pseudoscalar)  resonance $R$ with
 mass of about 60 GeV, coupling photons and $Z^{0}$-bosons only.

Proceeding from the requirement of the Lorentz and gauge
invariance, the most general form of the
$R\gamma\gamma$, $R\gamma Z$ and $RZZ$ vertices will read as follows:
\begin{eqnarray}
R^+ \gamma\gamma & : &  \frac{ g_{\gamma\gamma} }{M_{R}}
 (g^{\mu\nu}(k_{1}k_{2})-k_{1}^{\nu}k_{2}^{\mu})
 e_{1}^{\mu}e_{2}^{\nu},    \nonumber\\
R^{-}\gamma\gamma & : & \frac{g_{\gamma\gamma}}{M_{R}}
 \varepsilon^{\mu\nu\alpha\beta}k_{1}^{\mu}k_{2}^{\nu}
 e_{1}^{\alpha}e_{2}^{\beta},  \nonumber\\
R^\pm \gamma Z & : & \frac{ g_{\gamma z1} }{M_{R}}
 (g^{\mu\nu}(k_{1}k_{2})-k_{1}^{\nu}k_{2}^{\mu})
 e^{\mu}V^{\nu}+\frac{g_{\gamma z2}}{M_R}
 \varepsilon^{\mu\nu\alpha\beta}k_{1}^{\mu}k_{2}^{\nu}
 e^{\alpha}V^{\beta},  \nonumber\\
R^\pm ZZ & : & g_{zz1} M_{Z^0}  g^{\mu\nu} V_1^{\mu}V_2^{\nu}+
\frac{g_{zz2}}{M_R}k_1^{\nu}k_2^{\mu}V_1^{\mu}V_2^{\nu}+
\frac{g_{zz3}}{M_R} \varepsilon^{\mu\nu\alpha\beta}k_{1}^{\mu}k_{2}^{\nu}
 V_{1}^{\alpha}V_{2}^{\beta}, \nonumber
\end{eqnarray}
  where $R^+(R^-)$ denotes scalar (pseudoscalar) resonance;
$k_1$ and $k_2$ are momenta of two photons (photon and $Z$--boson);
 $e^{\nu}$
($V^{\nu}$) is photon ($Z^{0}$-boson) polarization vector.
 The factor $1/M_R$
 is introduced to render the coupling constants dimensionless.

    In [10,11] a number of experiments has been analysed,
in which hypothetic resonance $R$ can decay into $\gamma\gamma$. As a
 consequence the
following bounds on the couplings (in fact, on the values
$g\cdot \sqrt{Br(R\to \gamma\gamma)}$, see [10,11]) are obtained:
\begin{eqnarray}
&& g_{\gamma\gamma}\,\leq\,(2.92\pm 2.1)\cdot 10^{-3},
g_{\gamma z1}\,\leq\,(3.0\pm 6.8)\cdot 10^{-3},
g_{\gamma z2}\,\simeq\,0.\pm 0.02, \nonumber \\
&& g_{zz1}\,\leq\,0.220\pm 0.1, \qquad
| g_{zz2} | \,\leq\,0.571\pm 1.1 , \qquad
g_{zz3}\,\simeq\,0\pm 1.6.
\end{eqnarray}

    The processes of $R$ production
followed by the photon pair decay are considered. Thus
all cross-sections will also
depend on the couplings, multiplied by $\sqrt{Br(R\to\gamma\gamma)}$.
\section{\bf CROSS-SECTION OF THE {\boldmath$R$}--RESONANCE PRODUCTION}

    In this Section we consider the $R$--resonance production
in the $p\bar p$--collisions at the FNAL hadron collider
($\sqrt{s}\,=\,1.8$ TeV).
In the frames of the considered
model (see Section 1), the $R$--resonance production takes place
due to $q\bar q$ annihilation to the virtual $\gamma^{\ast}$ or
$Z^{\ast}$ (see Fig.1):
\be
q\bar q \,\to\, \gamma^{\ast} (Z^{\ast})\,\to\,l\bar l\,R\,
(\,\to\,\gamma\gamma)
\ee

    We concentrate our attention on the decays $Z^{\ast}$ (or
$\gamma^{\ast}$) to the lepton pair ($l\bar l\,=\,e^+e^-\,,\,
\mu^+\mu^-\,,\,\nu\bar\nu$) in the final state. It is obvious,
that $Z^{\ast}$ (or $\gamma^{\ast}$) decay to the $q\bar q$--pair
(i.e. process $q\bar q \,\to\, \gamma^{\ast} (Z^{\ast})\,\to\,q\bar q\,R\,
(\,\to\,\gamma\gamma)$) will be dominant. However, the existing QCD
background in this channel essentially exceeds the signal.
Thus we limit ourselves to considering that kind of processes:
\be
p\bar p\, \to \, l^+l^-\,R\,X\,\to\, l^+l^-\gamma\gamma\, X \\
p\bar p\, \to\, \nu\bar\nu \,R\,X\,\to\, \nu\bar\nu\gamma\gamma\, X
\ee

    In the framework of the parton model the cross-section of
processes (2) and (3) has the following form:
\begin{eqnarray}
&& \sigma( p\bar p \rightarrow l^{+}l^{-}(\nu\bar\nu)\gamma\gamma X)
 =  \nonumber \\
&& \sum_{q=u,d,s,c}
    \int dx_1 dx_2
\left[ f_{q}(x_1,Q^2) f_{q}(x_2,Q^2)+
 f_{\bar q}(x_1,Q^2) f_{\bar q}(x_2,Q^2)\right] \times \nonumber \\
&&
{}~~~~~~~~~\hat\sigma(q\bar q \rightarrow l^{+}l^{-}(\nu\bar\nu) R)
Br(R\rightarrow\gamma\gamma)\cdot N_l,
\end{eqnarray}
where $f(x,Q^2)$ are quark structure functions from [12],
the evolution parameter $Q^2$ is chosen to be equal
to the parton center-of-mass energy squared
$\hat s\equiv x_{1}x_{2}s$, and $N_l\,=\,1(3)$
for $l^+ l^-\,(\,\nu\bar\nu\,)$.

 Here we must note that the phenomenological model introduced in
the previous Section, being an effective one, eventually
violates unitarity, but the smallness of the couplings (and, probably, the
greatness of the scale of the new physics possibly involved) pushes up the
unitary
limit to energies as high as tens of TeV. Thus,
the model is still working at the FNAL collider energies.
 \begin{table}
 \begin{center}
 \begin{tabular}{|c||c|c|c|c|c|c|c|}
 \hline
  Cut&1&2&3&4&5&6&7 \\
 \hline \hline
 $P_{t~min}^{l^{+}(l^{-})},$ GeV$/c$& --- &5&10&15&5&10&15 \\
 \hline
 $P_{t~min}^{\gamma},$ GeV$/c$& --- &\multicolumn{6}{|c|}{10} \\
 \hline
 $|\eta^{\gamma}|_{max}$& --- &
  \multicolumn{3}{|c|}{3.0}&\multicolumn{3}{|c|}{1.5}\\
 \hline
 \end{tabular}\\
 \tablecaption{ The cuts used in the calculations.}
 \end{center}
 \end{table}

 Furthermore, to ensure good experimental efficiency
we require that
the photons be detected in the central $\eta$ interval and that they have
sufficient $P_{t}$, which compensates for the
lepton pair.
 The set of cuts on the phase space of final particles,
that we applied integrating (4), is summarized in Table 1
(naturally, only the final photons cuts were used in the
$\nu\bar\nu\gamma\gamma$ analysis).

    The total cross-sections of processes (2) and (3)
(kinematical cuts from Table 1 are taken into account) are presented
in Table 2.
\begin{table}
 \begin{center}
 \begin{tabular}{|c||c|c|c|c|c|c|c|}
 \hline
  Cut&1&2&3&4&5&6&7 \\
 \hline \hline
 $\sigma (p\bar p \rightarrow l^+l^-\gamma\gamma ~X)$, pb&
 0.076&0.068&0.067&0.065&0.062&0.061&0.060 \\
 \hline
 $\sigma (p\bar p \rightarrow \nu\bar\nu\gamma\gamma ~X)$, pb&0.42 &
  \multicolumn{3}{|c|}{0.39}&\multicolumn{3}{|c|}{0.35} \\
 \hline
 \end{tabular}\\
\tablecaption{ The cross sections of process (2).}
 \end{center}
\end{table}

     One can expect about 10 events in process (2) and
about 40 events in
process (3) for the total integrated luminosity
$\int {\cal L} dt\,=\,100$ pb$^{-1}$.

    The distributions over the various
variables for final leptons and photons from processes (2) and (3) are
presented on Fig.2 and 3.
As we can see from Fig.2(c) and 2(d) the final leptons
and photons are produced with
rather large transverse momenta ($<p_T^l>\,\simeq\,50$ GeV and
$<p_T^\gamma >\,\simeq\,30$ GeV).
It is necessary to note, that the final leptons and photons have a good
separation from each other (see Fig.3(d)). Thus, all the events of that kind
must contain photons with high transverse momenta $p_T$
and $m_{\gamma\gamma}\,=\,60$ GeV. This is the most characteristic
feature of these events.

    Hence, a pronounced topology of the investigated events will allow
the background of these events to be strongly suppressed and make
their investigation in the CDF and D0 detectors quite a real problem.
\bc
{\bf CONCLUSIONS.}
\ec
The limits that low energy $e^+e^-$
experiments set on the couplings of the hypothetical (pseudo)scalar
resonance $R$ with photons and $Z$--bosons from the LEP
low energy experiments data, still offer an opportunity
to investigate these events at FNAL hadron collider energies.
 The processes of the $R$--resonance production in
\bc
 $p\bar p\to l^+l^-~R~X\to l^+l^- \gamma\gamma ~X$ and \\
 $p\bar p\to \nu\bar\nu~R~X\to \not{\!\!E}_T \gamma\gamma ~X$\\
\ec
modes have the cross-sections of 0.05 -- 0.08 pb (depending on
experimental cuts)
and 0.3 - 0.5 pb, respectively. The high luminosity
($10^2$ pb$^{-1}$/year), supposed to be available at Fermilab
in the nearest future, will allow one to obtain a noticeable
number of events corresponding to these cross--sections.
 About 10 $l^+l^-\gamma\gamma$ and 40
$\nu\bar\nu\gamma\gamma$
events could be observed under these experimental conditions.

 The processes under study possess pronounced features, which make
observation of such events possible.

    Thus, the FNAL hadron collider gives a good opportunity
to check independently the model based on the LEP data. All data,
which will be obtained from CDF and D0 detectors may increase
the constraints, received from the LEP data, and allow one to draw a final
conclusion about the existence of that kind of resonance.
\vspace{0.5cm}

{\large\bf Acknowledgments.}\\
We are grateful to S.L.Linn, V.F. Obraztsov and A.M. Zaitsev  for stimulating
discussions. One of us (S.R.S.) would like also to thank Physics Department of
the Florida State University, where the work
was completed, for their kind hospitality and financial support.

\vspace{2cm}
\br
{\it Received June 16, 1994.}
\er

\newpage
\vbox{\vglue 2.5cm}
\bc
{\Large \bf Captions of figures.}
\ec
\begin{enumerate}
\item [Fig.1.]   The Feynman diagrams for the process $q \bar q \,\to\,
l\bar l \,R\,(\,\to\,\gamma\gamma)$ are presented.
\item [Fig.2.]   The differential cross-sections of processes (2) and
(3): a) -- the
invariance mass dilepton distribution; b) -- the transverse
momentum distribution for resonance; c) -- the transverse momentum
distribution of final leptons;
d) -- the transverse momentum distribution of final photons.
\item [Fig.3] The differential cross-sections of processes (2) and
(3): a) -- the resonance rapidity distribution;
b) -- the maximum rapidity distribution from the ones
both final leptons;
c) -- the maximum rapidity distribution from the ones
both final photons; d) --
the $\delta R$ distribution between leptons and photons,
where $\delta R\,=\,\sqrt{(\delta \phi)^2\,+\,(\delta \eta)^2}$.
\end{enumerate}

\newpage

\begin{figure}[h]
\begin{center}
\ForceWidth{\textwidth}
\BoxedEPSF{ df.ps }
\caption{}
\end{center}
\end{figure}

\begin{figure}[h]
\begin{center}
\ForceWidth{\textwidth}
\BoxedEPSF{ d1.ps }
\caption{}
\end{center}
\end{figure}

\begin{figure}[h]
\begin{center}
\ForceWidth{\textwidth}
\BoxedEPSF{ d2.ps }
\caption{}
\end{center}
\end{figure}


\newpage

\begin{thebibliography}{10}
\bibitem{L3}
  L3 Collab., O. Adriani et al., Phys. Lett. B 295 (1992) 337;\\
  S.C.C. Ting, Preprint CERN-PPE/93-34, 1993
\bibitem{ALEPH}
  ALEPH Collab., D. Busculic et al., Phys. Lett. B 308 (1993) 425
\bibitem{OPAL}
  OPAL Collab., P.D. Acton et al., Phys. Lett. B 311 (1993) 391
\bibitem{DELPHI}
  R. Henriques et al., DELPHI Note 92-170 PHYS 253, 1992;\\
  O. Barring and A. De Min, DELPHI Note 92-171 PHYS 254, 1992;
\bibitem{JW}
  S. Jadach and B.F.L. Ward, Phys. Lett. B 274 (1992) 470
\bibitem{MM}
  M. Martinez and R. Miquel, Phys. Lett. B 302 (1993) 108
\bibitem{Summers}
  D.J. Summers, Phys. Lett. B 302 (1993) 326
\bibitem{Italians}
  A. Ballestrero, E. Maina and S. Moretti, Phys. Lett. B 305 (1993) 312
\bibitem{other models}
Barger V. et al,  Preprint ANL--HEP--PR--92--102, and
MAD/PH/728, University of Wisconsin, 1992.

Bando M.,Maekawa N.,  Preprint KUNS 1174 HE(TH) 92/14, Kyoto University, 1992.

Cvetic G., Nowakowski M., Yue--Liang Wu,  Preprint DO--TH/92--24, Dortmund
 University, 1992.

Garisto R. and Ng J.N.,  Preprint TRI--PP--92--124, University of British
Columbia, Vancouver, 1992.

Geng C.Q., Whisnant K., Young B.-L.,  Preprint INS--J 4917, Institut for
Nuclear Study, Tokyo , 1992.

Kang K., Knowles I.G., White A.R.,  Preprint ANL--HEP--PR--93--4, Argonne
National Laboratory, Argonne, 1993.

Lubicz V.,  Preprint ROM--P 925, Roma University, Italy, 1992.
\bibitem{The Foundations 1}
  V. Litvin and S. Slabospitsky, Yad. Fiz. v.57(5), 877, 1994.
\bibitem{The Foundations 2}
  V. Litvin and S. Slabospitsky, Preprint IHEP 94--22, 1993.
\bibitem{CTEQ}
  J. Botts, J.G. Morfin, J.F. Owens, J. Qiu, W.K. Tung and H. Weerts,
  MSUHEP-92-27, Fermilab-Pub-92/371, FSU-HEP-92-1225, ISU-NP-92-17
\bibitem{CDF descr}
  CDF Collab., F. Abe et al., NIM A271 (1988) 387  \\
  D0 Collab., E. Adachi et al., NIM A338 (1994) 185
\end{thebibliography}
\end{document}